\documentclass[longauth]{aa} 

\usepackage{graphicx}

\usepackage{latexsym} 
\usepackage{graphicx} 
\usepackage{subfigure} 
\usepackage{longtable} 
\usepackage{amsmath} 
\usepackage{supertabular} 
\usepackage{natbib} 
\usepackage[latin1]{inputenc} 
\usepackage[thinspace]{SIunits} 

\begin{document} 

\addunit{\parsec}{pc} 
\addunit{\ergios}{ergs} 
\addunit{\anho}{yr} 
\addunit{\edad}{\giga\anho} 
\def \cm {\centi\meter} 
\def \be {\begin{equation}} 
\def \ee {\end{equation}} 
\def \vel {\kilo\meter\per\second} 
\def \phu {\kilo\meter\,\reciprocal\second\mega\reciprocal\parsec} 
\def \kev {\kilo\electronvolt} 
\def \dt {\gram\,\rpcubic\cm} 
\def \s   {\second} 
\def \bd  {$\ldots$} 
\def \cd  {$\cdots$} 
\def \sr {$\pm$} 
\def \om {$\Omega_{\textrm{m}}$} 
\def \omm {\Omega_{\textrm{m}}} 
\def \ox  {$\Omega_{\textrm{x}}$} 
\def \oxx  {\Omega_{\textrm{x}}} 
\def \siom {$\sigma_{\Omega_{m}}$} 
\def \siox {$\sigma_{\Omega_{x}}$} 
\def \si {$\sigma_{n}$} 
\def \ho {$H_0$} 
\def \epk {$E_{\textrm{peak}}$} 
\def \epkk {E_{\textrm{peak}}} 
\def \pbo {$P_{\textrm{bolo}}$} 
\def \sbo {$S_{\textrm{bolo}}$} 
\def \siho {$\sigma_{H_0}$} 
\def \sith {$\sigma_{\theta}$} 
\def \rcmb {${\cal R}$} 
\def \bao {${\cal A}$} 
\def \chidof {$\chi^2_{\text{dof}}$} 
\def \zc {$z_{\text{card}}$} 
\def \zac {$z_{\text{acc}}$} 
\def \te {\theta} 
\def \sg {\sigma} 
\def \m {\text{m}} 
\def \trt {\tau_{\text{\tiny RT}}} 
\def \tlag {\tau_{\text{\tiny lag}}} 
\def \prh {Prior $H_0$} 
\def \eg {E_{\gamma}}

\title{Hubble diagram of gamma-ray bursts calibrated with Gurzadyan-Xue 
cosmology} 

\author{H. J. Mosquera Cuesta\inst{1} \and R. Turcati\inst{1} \and C. Furlanetto \inst{1} \and H. G. Khachatryan \inst{2} \and \\  S. Mirzoyan\inst{2} \and G. Yegorian \inst{2} \\ 
\inst{1} Instituto de Cosmologia, Relatividade e Astrof\'isica(ICRA-BR), Centro Brasileiro de Pesquisas F\'isicas. \\ \hskip 0.3truecm Rua Dr. Xavier Sigaud 150, CEP  22290-180, Urca, Rio de Janeiro, RJ, Brasil \\ 
\inst{2} Yerevan State University and Yerevan Physics Institute, Yerevan, Armenia} 


\abstract{}{}{}{}{} 

\abstract
{Gamma-ray bursts (GRBs) being the most luminous among known cosmic objects 
carry an essential potential for cosmological studies if properly used as standard 
candles.}
{ In this paper we test with GRBs the cosmological predictions of the Gurzadyan-Xue 
(GX) model of dark energy, a novel theory that predicts, without any free parameters, 
the current vacuum fluctuation energy density close to the value inferred from the SNIa 
observations. 
We also compare the GX results with those predicted by the concordance scenario $\Lambda$-CDM.
} 
{According to the statistical approach by Schaefer (2007), the use of several empirical 
relations obtained from GRBs observables, after a consistent calibration for a specific 
model, enables one to probe current cosmological models. { Based on this  recently 
introduced method, we use the 69 GRBs sample collected by Schaefer (2007); and the most 
recently released SWIFT satellite data (Sakamoto et al. 2007) together with the 41 GRBs 
sample collected by Rizzuto et al. (2007), which has the more firmly determined redshifts. 
Both data samples span a distance scale up to redshift about 7. } } 
{We show that the GX models are compatible with the Hubble diagram of the Schaefer 
(2007) 69 GRBs sample. Such adjustment is almost identical to the one for the concordance 
$\Lambda$-CDM. 
>From this particular analysis we can obtain the corresponding values of the matter density 
parameter $\Omega_m$ describing GX models. { When the similar procedure is 
applied to the Rizzuto et al. (2007) and Sakamoto et al. (2007) SWIFT satellite data, 
we verify that the SWIFT sample does not delineate a Hubble diagram as clearly as featured 
by the 69 GRBs sample. } 
} 
{ { The analysis of the samples of Schaefer (2007) and those by Rizzuto et al. (2007) 
and of SWIFT (Sakamoto et al. 2007) shows that more data 
and efforts are needed to elucidate both issues}: the gamma-ray bursts/standard-candle and lack 
of a theoretical understanding of the physics subjacent to the empirical relations.}
{} 

\date{Received (\today)} 
\titlerunning{Hubble Diagram of Gamma-Rays Bursts} 

\authorrunning{Herman J. Mosquera Cuesta et. al.} 


\maketitle 

\section{Introduction}\label{int} 

Recently Gamma-Ray Bursts (GRBs) were used for cosmography aims, for the 
analyses of the 
high-redshift behavior of the $\Lambda$-CDM cosmological scenario, as well as 
for several alternative cosmologies (see \cite{Schaefer, Schaefer1, 
Bloom, Dai, Ghirlanda, Friedman, Liang, Liang1, Xu, Wang, Firmani}; and other 
papers). The potential power of GRBs for cosmological studies, obviously, 
resides both in their very high luminosity, which the highest among known 
astrophysical objects, and 
the fact that they undergo practically no extinction over cosmological 
distances. Hence, one can get the possibility to trace significantly higher 
distance scales than 
it is possible via 
supernovae, that is, to trace very deep in the expansion history of the universe. 
The situation however, not as simple.

Schaefer (2007) has developed a statistical approach based on the 
empirical correlations 
obtained from several observed GRBs characteristics and obtained the Hubble 
diagram for the GRBs after 
calibrated them for the concordance and other cosmological models. This approach 
has been used also in 
(\cite{Cuesta}). The key issue is the use of the empirical relations, e.g. of Ghirlanda,
Liang-Zhang and others, in the absence 
of understanding of their underlying nature, i. e., the genuine character of the 
scatter in the GRBs 
luminosity vs. luminosity indicators relations and of their mutual links. To 
this difficulty Schaefer (2007) has shown that, although the scatter of each of
empirical relations can be not as small, their joint action can lead to a smaller
scatter, useful for probing certain cosmological models. 
Increase of the statistics and deeper studies of the systematics and 
selection effects will certainly increase the informativity of Schaefer's approach. 

In the present paper we use the 
same approach as Schaefer (2007), to obtain the gamma-ray burst Hubble diagram for 
the cosmological models 
proposed by Gurzadyan and Xue (2003). The original motivation for GX 
cosmological models is the fact 
that, they predict the current vacuum fluctuation energy density close to the 
value inferred from 
the SNIa observations (\cite{Perlmutter,Riess1,Perlmutter1,Riess2,Riess3}) 
without any free parameter. 

\section{The Cosmological Models} 

Gurzadyan and Xue (2002, 2003) have derived a formula for the dark energy, which 
fits the observed value without free parameters 

\begin{equation} 
\rho_{GX}=\frac{\pi}{8}\frac{\hbar c}{L_{p}^{2}}\frac{1}{a^{2}}=\frac{\pi}%
{8}\frac{c^{4}}{G}\frac{1}{a^{2}}, \label{rhoLambda}%
\end{equation} 
where $\hbar$ is the Planck constant, the Planck length is $L_{p}=\left( 
\hbar G\right)^{\frac{1}{2}}c^{-3/2}$, $c$ is the speed of light, and $G$ is the 
gravitational 
constant. Here $a$ is the upper cutoff scale in computation of vacuum 
fluctuations and has to be 
close to the event horizon (\cite{DG}). According to (\cite{Zel67}), the 
vacuum energy 
(\ref{rhoLambda}) corresponds to the cosmological term.%
GX formula (\ref{rhoLambda}) defines a broad set of cosmological models 
(\cite{Ver06}). For the latter the existence of a separatrix was shown 
(\cite{Ver06a}), which divides the space of cosmological solutions into two 
classes:\ Friedmannian-like with initial singularity and non-Friedmannian 
solutions which begin with nonzero scale factor and vanishing matter density. 
Each solution is characterized by the single quantity, a density parameter 
which is defined in the same way as in the standard cosmological model 
$\Omega_{m}=\frac{8\pi G_{0}\mu_{0}}{3H_{0}^{2}},$ where $\mu$ is the matter 
density, $H$ is the Hubble 
parameter, and index "$0$" refers to their values today. The separatrix is given 
by 

\be 
\Omega_{sep}=\frac{2}{3}\frac{1}{1-\frac{K}{\pi^{2}}}\approx\frac{2}{3}, 
\label{sep} 
\ee 

where $K=\pm1,0$ parametrize the spatial curvature. The origin of the 
separatrix was revealed in (\cite{Khach07}), and attributed to existence of 
invariants in GX models. 

Analytical solutions for the GX models both for matter density and the scale 
factor are obtained (\cite{Ver06b}). It turns out that the most simple 
solutions for the scale factor are again those of the separatrix. In one model it 
is exponential, in the others they are polynomials. Vereshchagin and Yegorian 
(2006b, 2008) generalized GX models to include radiation and looked for
other consequences of the models.

The predictions of the GX models were shown to be compatible to supernovae and
Cosmic Microwave Background data in 
(\cite{DG}). A likelihood analysis of supernovae and radio galaxies 
data was performed in (\cite{Ver06c}, \cite{Khach07a}).

\section{Luminosity Distance Formula for GX-models} 
The models are described by the two equations for the mass density and scale 
factor 
(\cite{Ver06b}) 

\begin{gather} 
\dot{\mu}+3H\left(  \mu+\frac{p}{c^{2}}\right)  =-\dot{\mu}_{\Lambda}+\left( 
\mu+\mu_{\Lambda}\right)  \left(  \frac{2\dot{c}}{c}-\frac{\dot{G}}{G}\right) 
,\nonumber\\ 
H^{2}+\frac{kc^{2}}{a^{2}}-\frac{\Lambda}{3}=\frac{8\pi G}{3}\mu,\label{MD.1} 
\end{gather} 
where a dot denotes time derivative. 
For matter (pressure $p=0$) we have 
solution for matter mass density with GX-dark energy (\cite{Khach07}) 
\begin{equation} 
\mu_{m}(t)=\left(  b_{m}^{GX}+\frac{\pi a(t)}{4}\right)  \frac{c^{2}%
(t)}{G(t)a^{3}(t)},\label{MD.3} 
\end{equation} 
where $b_{m}$ is a GX-invariant for matter. For the scale factor 
\begin{equation} 
\dot{a}(t)=c(t)\sqrt{\frac{8\pi b_{m}^{GX}}{3a(t)}+\pi^{2}-k}.\label{MD.4}%
\end{equation} 
The luminosity distance $d_{L}$ is (\cite{Peebles,Daly}) 
\begin{gather*} 
d_{L}(z)=a_{0}f_{k}(\kappa_{s})(1+z),\\ 
\kappa_{s}=\frac{1}{a_{0}H_{0}}\int_{0}^{z}\frac{c(\acute{z})}{h(\acute{z}%
)}d\acute{z},\\ 
h(z)=\frac{H(z)}{H_{0}},1+z=\frac{a_{0}}{a} 
\end{gather*} 

where $\kappa_{s},z$ are normalized distance and redshift, respectively. 
The function $f_{k}(\kappa_{s})$ is defined as 
\begin{equation} 
f_{k}(x)=\left\{ 
\begin{array} 
[c]{c} 
\sin(x),k=1,\\ 
x,k=0,\\ 
\sinh(x),k=-1, 
\end{array} 
\right. 
\end{equation} 
here $k$ is the effective curvature $K-\pi^2/3$ as in (\ref{sep}). 
The luminosity distance $d_{L}(z)$ for GX models 
\begin{eqnarray} 
d_{L}(z)=a_{0}(1+z)f_{k} \left(\frac{1}{\sqrt{\beta}}\ln\left|\frac{g(z)-1} 
{g(0)-1}\,\,\frac{g(0)+1}{g(z)+1}\right| \right),\label{dlz} 
\end{eqnarray} 
where 
\begin{eqnarray} 
g(z)=\sqrt{\frac{\alpha}{\beta}(z+1)+1}, \hskip 0.3 truecm  \alpha = \frac{8 \pi 
b_{m}^{GX}}{3 a_0} \; , 
\hskip 0.3 truecm \beta = \pi^2 - k\; . 
\end{eqnarray} 
For the separatrix $\alpha=b_{m}^{GX}=0$ we have a simple equation for the 
luminosity 
distance 
\begin{equation} 
d_{L}(z)=a_{0}(1+z)f_{k} 
\left(\frac{\ln|z+1|}{\sqrt{\pi^{2}-k}}\right).\label{dlzsep} 
\end{equation} 

\section{Gamma Ray Bursts calibrated with GX models} 

For the present analysis we benefit of the largest GRBs sample having 
properly determined redshifts and luminosities currently available. The
first was constructed by Schaefer (2007) and includes 69 GRBs whose main 
observables: time lag, variability, peak energy, maximum energy in 
$\gamma$-rays and rise time, which were obtained from the GRBs data provided 
by many $\gamma$-ray and X-ray satellites, and ground-based observatories, 
as collected in Greiner's homepage: http://www.mpe.mpg.de/~jcg/grbgen.html 
{ The second one is the 41 GRBs sample with firmly determined redshift 
as collected by Rizzuto et al. (2007) from the SWIFT satellite. }

\begin{figure*}[hbt] 
{\vspace{0.75truecm}
\subfigure[\label{cali1} Time lag vs. $L$ ]{ 
\includegraphics[scale=0.35]{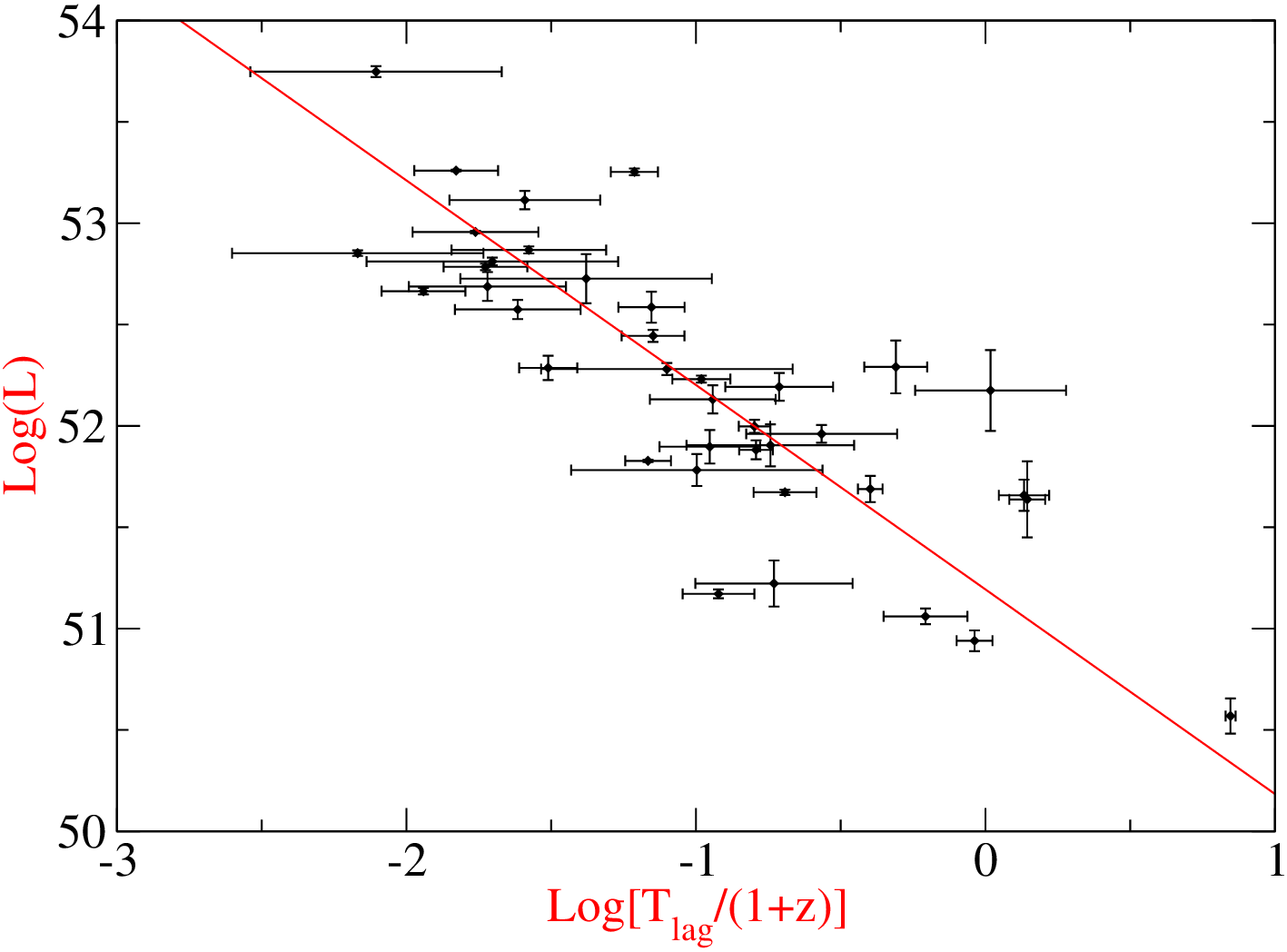}} 
\hspace{0.25cm} 
\subfigure[\label{cali2} Variability vs. $L$]{ 
\includegraphics[scale=0.35]{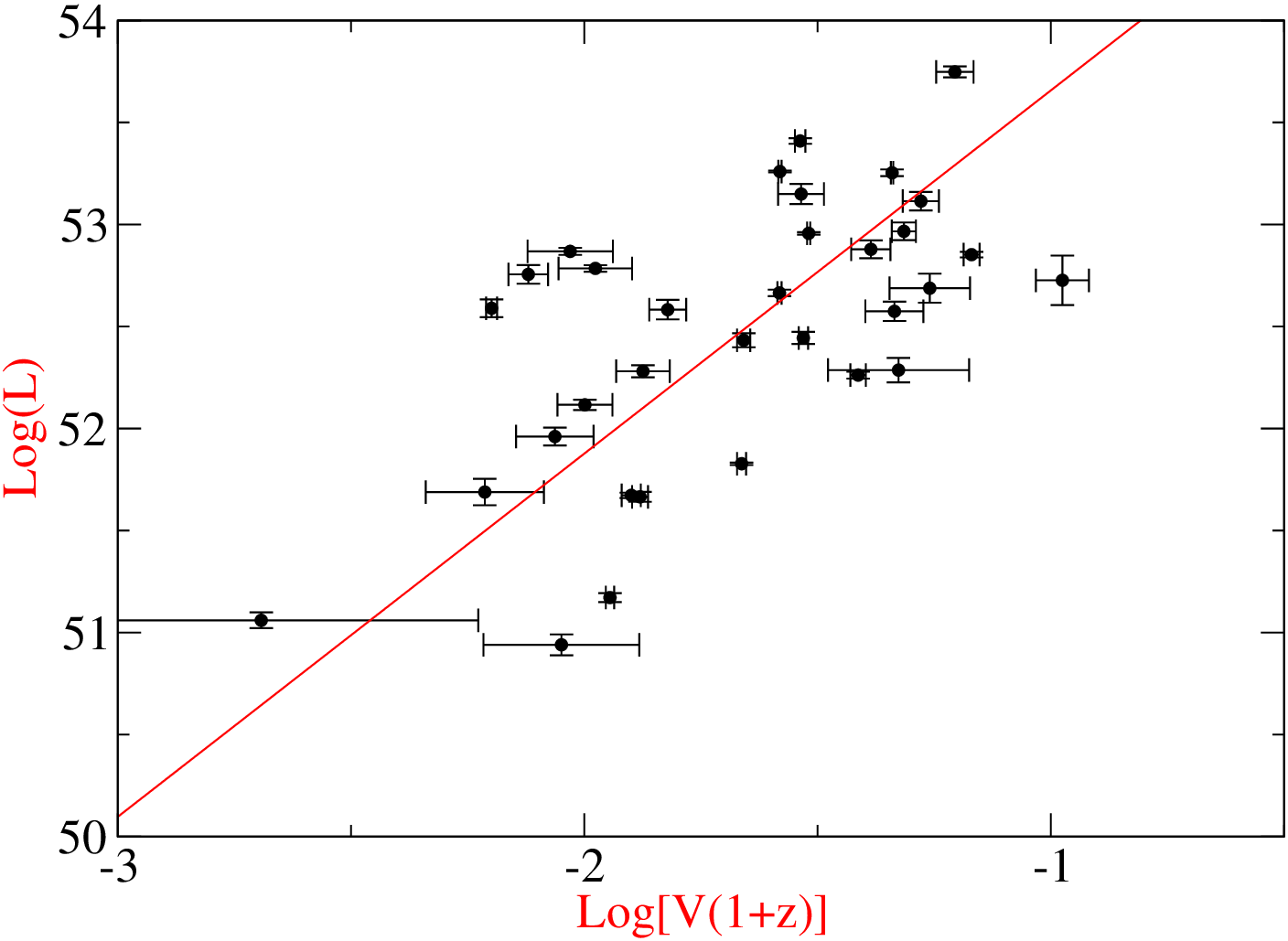}} }
\vspace{0.75truecm}{
\subfigure[\label{cali3} \epk \, vs. $L$]{ 
\includegraphics[scale=0.35]{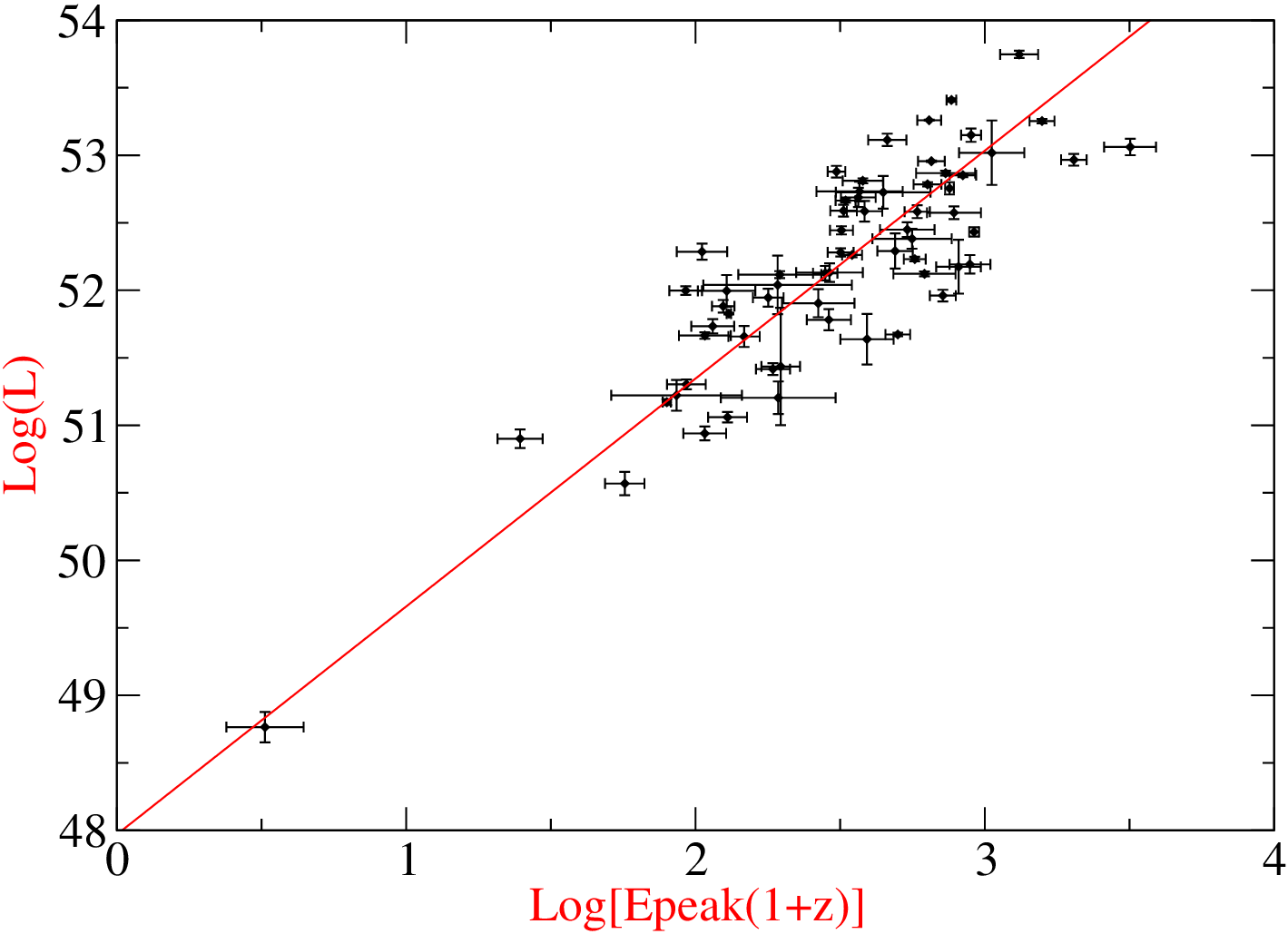} } 
\hspace{0.25cm} 
\subfigure[\label{cali4} \epk \, vs. E$_{\gamma}$\, (Ghirlanda Relation) ]{ 
\includegraphics[scale=0.35]{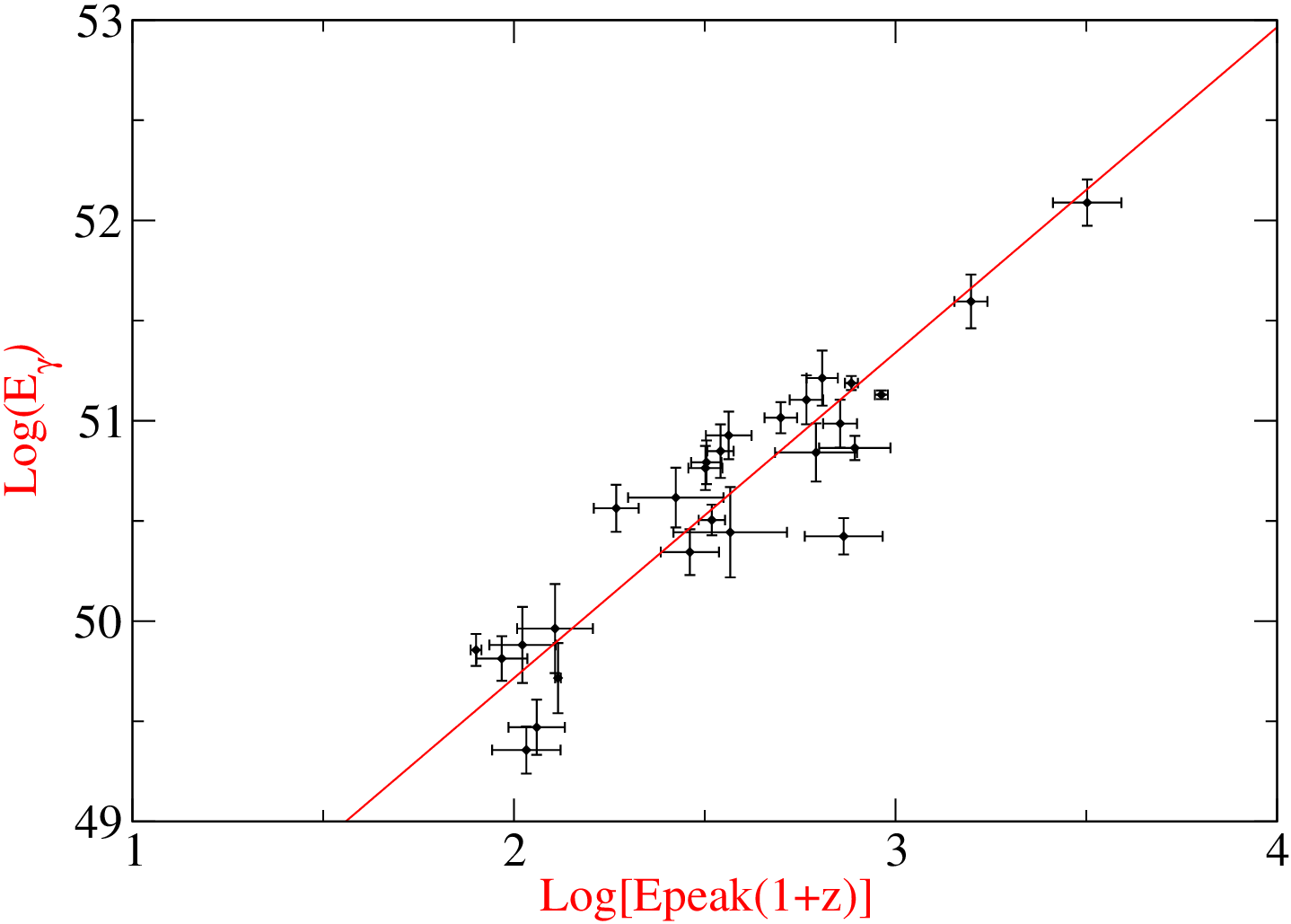} } }
\vspace{0.85cm}{ 
\subfigure[\label{cali5} Rise time vs. $L$ ]{ 
\includegraphics[scale=0.35]{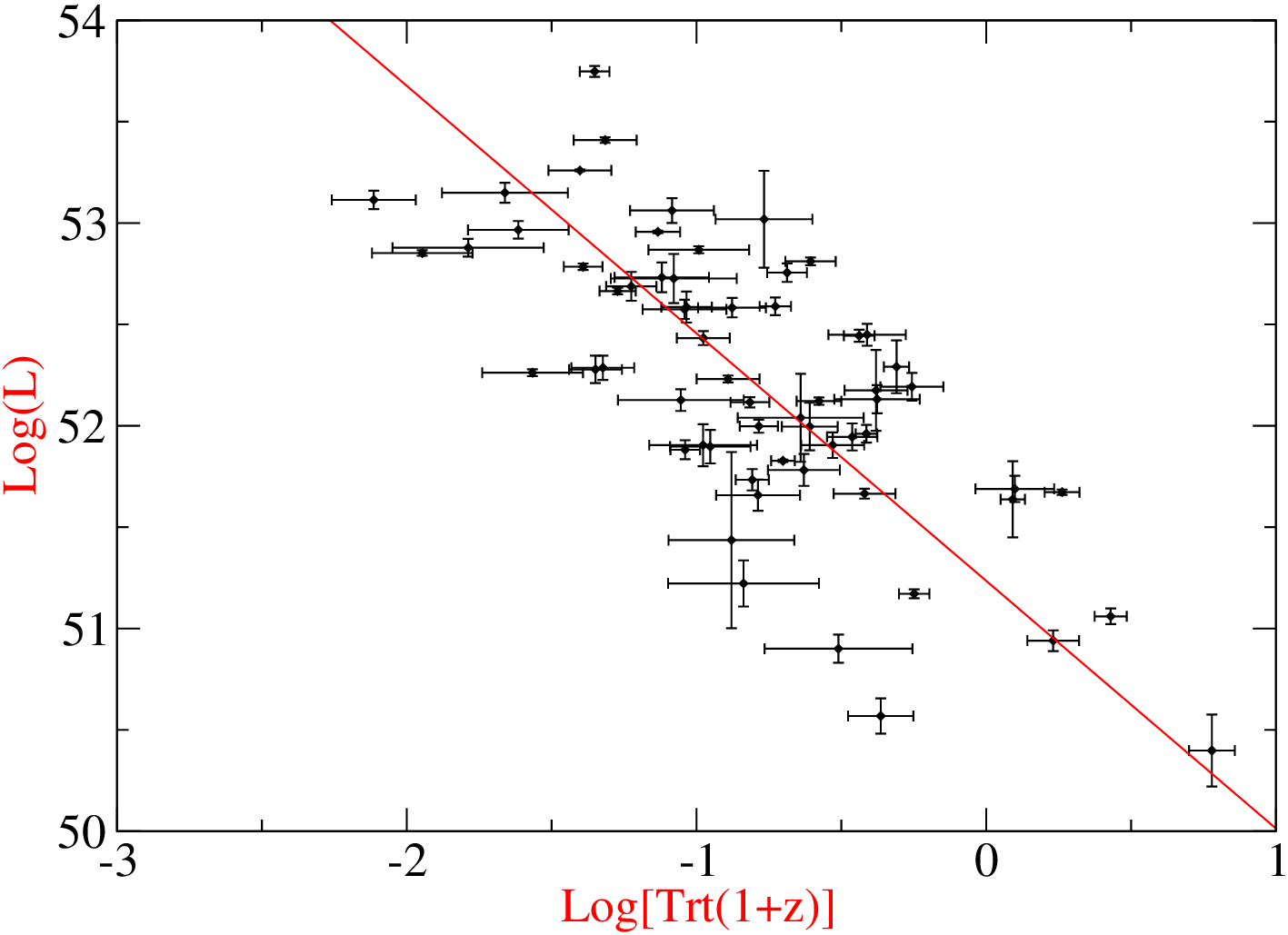} } } 
\caption{\label{calibrations} COLOR-ONLINE Results of the calibration procedure. 
Notice that all relations were corrected to the rest frame of the GRB and also 
by using the luminosity best-fit line obtained from the nonlinear regression method. 
(a) Time lags for 39 GRBs. 
(b) Variability for 51 GRBs vs. isotropic luminosity. 
(c) \epk\, values for 64 GRBs vs. isotropic luminosity. 
(d) \epk \,values for 27 GRBs vs. total burst energy in the gamma rays. 
(e) Rise time for 62 GRBs vs. isotropic luminosity.} 
\end{figure*}

{ 
We performed the calibration procedure (linear regression analysis) 
of five luminosity relations: time lag vs. luminosity ($\tlag-L$), 
variability vs. luminosity ($V-L$), peak energy vs. luminosity ($\epkk-L$), 
peak energy vs. geometrically corrected gamma-ray  
energy ($\epkk-\eg$), and risetime vs. 
luminosity ($\trt-L$) (\cite{Schaefer1,Cuesta}). We use the OLS Bisector 
method (\cite{isobe1}) to find a relation between each pair of these GRBs 
observational properties. The best-fit line for all luminosity 
relations is given by the general expression $\log{\cal L}= 
a+b\log{{\cal I}},$ where $\cal L$ is the luminosity and $\cal I$ is 
the luminosity indicator, and $a$ is the intercept and $b$ is the 
slope in each of the calibration plots here presented. Their 
uncertainties are: 

\be 
\sigma^2_{{\log(L_i)}} = \sigma^2_a + (\sigma_b \,x_i)^2 + (b\, \sigma_{x_i})^2 
+ \sigma^2_{\text{\tiny sys}} \; , 
\label{sdy1}                                                                
\ee 

where $\sigma_{\log{L_i}}$ defines the standard deviation in the 
luminosity ${L_i}$, $\sigma_a$ is the standard deviation in the 
intercept $a$, $\sigma_b$ is the standard deviation in the intercept 
$b$, $\sigma_{x_i}$ is the standard deviation in each $x_i$ variable 
representing $\log{I}$, and $\sigma_{\text{\tiny sys}}$ is the 
systematic error associated to each luminosity ${L_i}$ estimate. 
The results of the calibration procedure are given in the 
Table~\ref{tabrelac}, and the plot for all luminosity relations is 
given in the Fig.~(\ref{calibrations}). 

\begin{table}[hbt] 
 \caption{\label{tabrelac}Calibration Results} 
  \begin{tabular}{l c c c  c c } 
 Luminosity Relation        & $a$   & $\sigma_a$  & $b$   & $\sigma_b$ & 
$\sigma_{\text{\tiny sys}}$\\ 
                            &       &             &       &            &        
                    \\ 
 \hline 
 $\tau_{\text{lag}}-L$      & 52.20 &  0.07       & -1.01 &  0.09  &  0.36  \\ 
 $V-L$                      & 52.41 &  0.08       &  1.78 &  0.19  &  0.47  \\ 
 \epk-$L$                   & 52.15 &  0.05       &  1.69 &  0.10  &  0.41  \\ 
 \epk-E$_{\gamma}$          & 50.49 &  0.05       &  1.62 &  0.11  &  0.21  \\ 
 $\tau_{\text{\tiny RT}}-L$ & 52.45 &  0.07       & -1.22 &  0.11  &  0.47  \\ 
 \hline 
\end{tabular} 
 \end{table} 
 
}

Using the same method as Schaefer (2007), we obtained the best estimated 
distance moduli $\bar{\mu}_i$\footnote{For an outlying source of apparent; 
$m$, and absolute; $M$, magnitudes, distance estimates are made through the 
{\sl distance-modulus}: $\bar{\mu} \equiv m - M$, which is related to the 
luminosity distance $d_L$ (given below in units of Mpc) 

\be 
d_L = a_0 (1 + z) \int_a^{a_0} \frac{da}{a \dot{a}} \; , 
\label{lum-distance} \ee 
through the expression 
\be 
\bar{\mu}({z}) \equiv m - M = 5 \log_{10} d_L ({z}) + 25 \, . 
\label{distance-redshift} 
\ee 
} 

\begin{equation} 
\label{muto} 
\bar{\mu}  =  \frac{1}{w}\sum_{i=1}^5 w_i \bar{\mu}_i, \quad  w_i=1/\sigma^2_i 
\quad 
\textrm{and}\quad  \left[w=\sum_{i=1}^5 w_i\right] \; , 
\end{equation} 

where the summation is over the relations with available data, $\bar{\mu}_i$ is 
the best estimated distance modulus from the $i$-th relation, and $\sigma_i$ is 
the corresponding uncertainty. Then applying the error propagation law to the 
equation (\ref{muto}) we obtained the standard deviation associated to this best 
estimated as $\sigma_{\bar{\mu}}=1/\sqrt{w}$. 

\indent The Fig.(\ref{HD-GX-MODEL}) presents the Hubble Diagram for 
the 69 GRBs calibrated with the GX models and its comparison with 
the $\Lambda$-CDM model, for the parameters given in the inset. 
Fig.(\ref{dh}) shows the Hubble Diagram for the 69 GRBs calibrated 
with the GX models obtained after a slight variation of the 
separatrix solution for each cosmological model, i.e., depending on 
the curvature $k = 0, -1$. 

\begin{figure*}[hbt] 
\begin{center} 
\vspace{0.5cm} 
\subfigure[\label{ks=0} k = 0  ]{ 
\includegraphics[scale=0.325]{Figure3_k=0.eps}} 
\hspace{0.5cm} 
\subfigure[\label{ks=-1} k = -1]{ 
\includegraphics[scale=0.325]{Figure3_k=-1.eps}} 
\caption{\label{dh} COLOR-ONLINE  Hubble Diagram of 69 GRBs 
calibrated with GX models. (a) Case $k=0$ and different values of 
$\omm$. (b) Case $k=-1$ and different values of $\omm$. The 
corresponding parameters are indicated in the insets.} 
\end{center} 
\end{figure*} 


\begin{figure}[hbt] 
\begin{center} 
\vspace{0.75cm}
\includegraphics[scale=0.345]{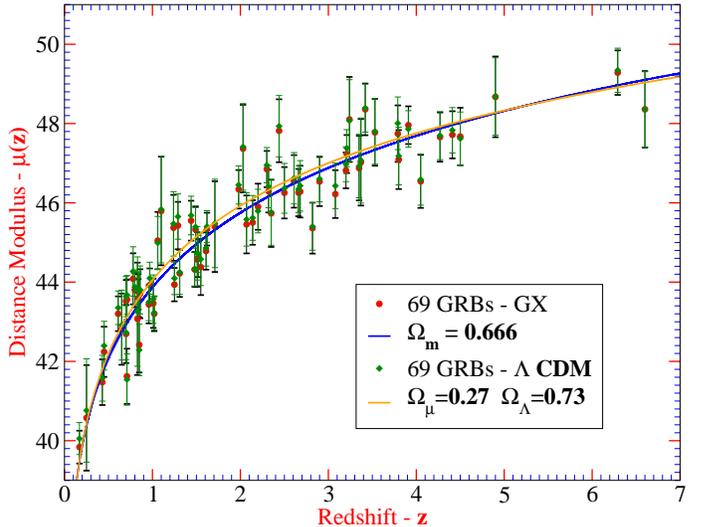}
\caption{\label{HD-GX-MODEL} COLOR-ONLINE  Hubble diagram of 69 GRBs 
calibrated with GX models as obtained from the empirical relations 
provided by Fig.-\ref{calibrations}. The blue line represents GX models 
calibrated with $\omm=0.66$, \,$k=0$ and $H_0=70$\, \phu. Orange line is 
for the concordance cosmology $\omm=0.27$, $\Omega_{\Lambda} = 0.73$. } 
\end{center} 
\end{figure}

\section{Results} 

\subsection{GX compatibility with the 69 GRBs sample }

\begin{figure*}[hbt] 
\begin{center} 
\subfigure[\label{k=0} k = 0]{
\includegraphics[scale=0.325]{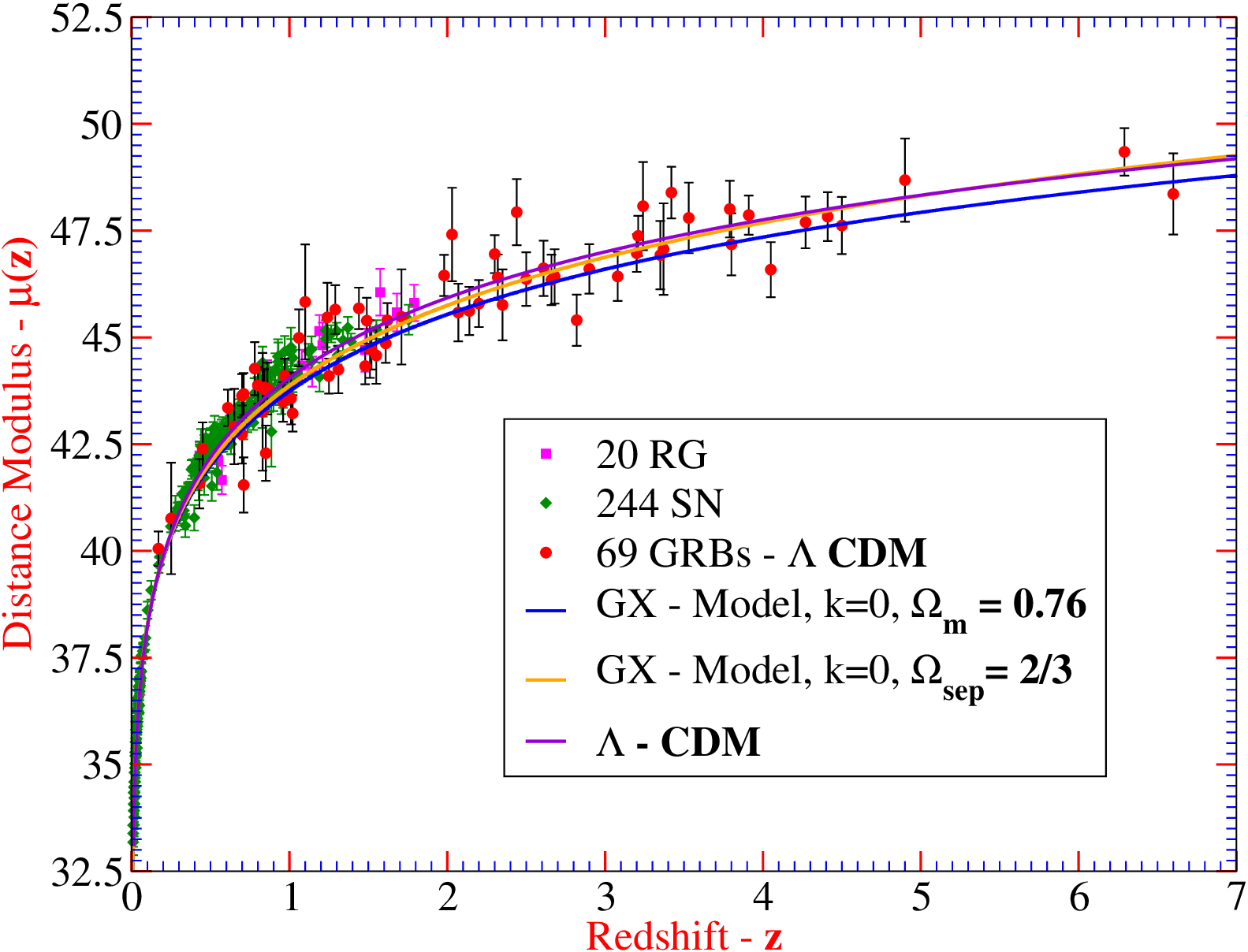}} 
\hskip 0.5truecm 
\subfigure[\label{k=-1} k = -1]{
 \includegraphics[scale=0.325,clip]{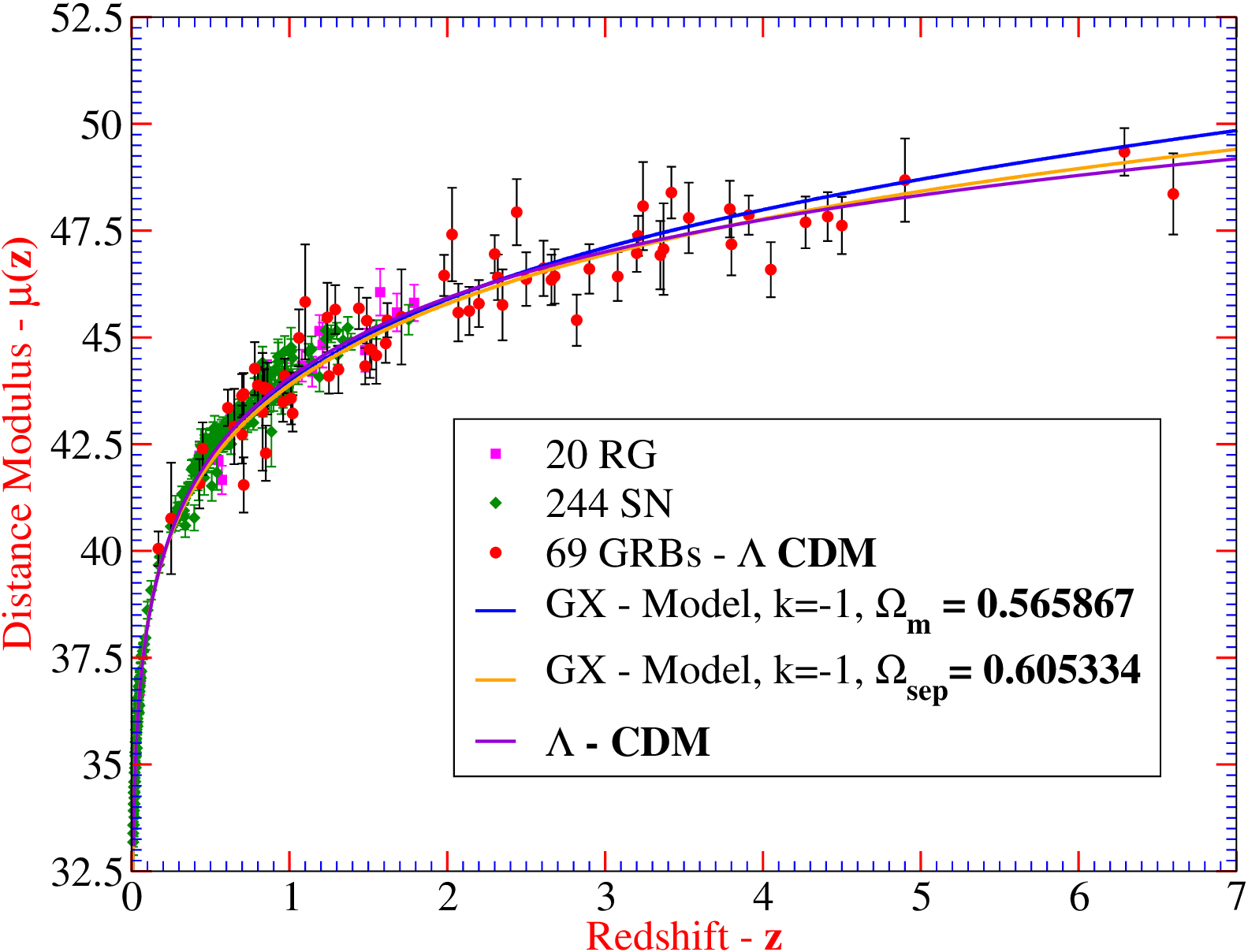}}
\caption{COLOR-ONLINE  a) The Hubble diagram for SNIa, radio-galaxies (RG), and GRBs calibrated 
both with GX model with $k=0$, $H_0 = 70$\, \phu, and $\Lambda$-CDM with $H_0 = 73$\, \phu. 
SNIa: green points. RG: magenta points. GRBS: red points and black error bars. b) Hubble diagram 
for GX with $k=-1$ and $H_0 = 70$\, \phu} 
\end{center} 
\label{k0} 
\end{figure*}

\begin{figure}[ht] 
\begin{center} 
\includegraphics[width=\hsize,clip]{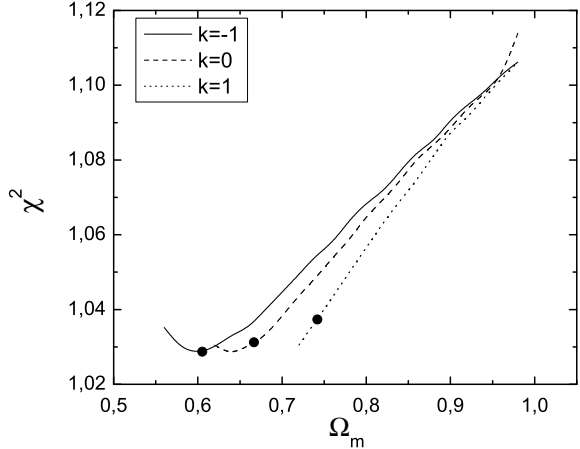} 
\caption{COLOR-ONLINE  $\chi^2$ for the Hubble diagrams vs the 
$\Omega_m$ for $k=0,-1,1$ GX-models. The fit points (circles) 
correspond to the separatrix.} 
\label{chi} 
\end{center} 
\end{figure}

We now compare GX models with the data set of 69 GRBs of 
\cite{Schaefer1}. We use general least square technique for all 
models with different density and curvature parameters. We find the 
best fit as $\Omega_m=2/3,\,\, k=0$. The $\chi^2$ value for that 
parameters is $1.037$ with 68 DoF. { The best fit curve is shown 
in Fig.(\ref{k=0}) for the value of the curvature parameter $k=0$, 
while in Fig.(\ref{k=-1}) the best fit curve is shown for $k=-1$.
In both figures, which combine gamma-ray bursts, supernova type Ia 
and radio-galaxies, several other HD from the GX model are plotted for 
different parameter $\Omega_m$, as indicated. Notice that the discrete 
set of data in the curves of Fig.(\ref{chi}) are due to the lack of an 
extensive study in the parameter space of $\Omega_m$. We plan to address 
this point in a forthcoming communication. }


For the same parameters we have $\chi^{2}\approx 2$ with 264 DoF for 
SN (\cite{Khach07a}). The pair best fit values of the matter density 
parameter are close to the separatrix value for $k=0,-1,1$, which once 
again shows the important role of the separatrix in GX models. 
Indeed, we see that the separatrix fits the best the observational 
data, with the shown values of the matter density parameter 
$\Omega_m$ and the curvature $k$. 



Concerning the $k=-1$ models, let us note that the ellipticity detected in the 
Cosmic Microwave Background radiation temperature maps is characteristic of 
photon beam motion in hyperbolic spaces (Gurzadyan et al 2005, 2007). Thus, 
it can act as a model-independent indication of the non precisely zero curvature 
of the Universe (Penrose 2005, Wiltshire 2007). 

{ 
\subsection{GX compatibility with the 41 SWIFT Satellite GRBs sample }

In this section we use the recently released SWIFT satellite, BAT instrument GRBs data (Sakamoto 
et al. 2007), namely, a sample of 41 GRBs known to have firmly 
determined redshifts (Rizzuto et al. 2007). We performed a similar procedure as Schaefer (2007), 
by constructing only three (3) luminosity indicator empirical relations from the SWIFT data. In
particular by using the two variabilities defined by Rizzuto et al. (2007), which were constructed 
taking into account the specific operational characteristics of the BAT instrument when analyzing 
the GRBs mask-tagged light curves. 
Those luminosity vs. luminosity indicator empirical relations are as follows [see a more detailed 
discussion, definitions and references in Rizzuto et al. (2007)]: 

\begin{itemize}
\item{Luminosity vs. Indicator: $V_{\rm R}$ Variability.---} We built this relation based on 
data of $V_{\rm R}$ variability provided by Rizzuto et al. (2007) and SWIFT Flux data from 
Sakamoto et al. (2007). Rizzuto et al. (2007) proved that a Poissonian variance describes the 
statistical fluctuations of the GRBs light curves (see Apendix A in Rizzuto et al. 2007). The 
main reason is that the count rates that were used in the analysis already had their background 
substracted 
\item{Luminosity vs. Indicator: Peak Luminosity.---} This relation was constructed for each GRBs 
as given in Table 1. of Rizzuto et al. (2007), which was obtained after extracting the mask-tagged 
light curve by using a binning time of 50 ms in the energy band [15-350] KeV (redshift corrected). 
The 1 s time interval with the highest total counts was found, and assumed as the time interval 
corresponding to 1 s peak count rate in the subsequent analysis 
\item{Luminosity vs. Indicator: $V_{\rm LP}$ Variability.---} The same procedure as above, but 
notice that it was supposed that no extra-Poissonian variance had to be substracted from the 
already mask-tagged light curves
\end{itemize}

Our results for the SWIFT GRBs data analysis are presented in the Figure-\ref{swift-calibrations},
where the three empirical relations (6(a) $V_{\rm LP}$ vs. $L$, 6(b) $V_{\rm R}$ vs. $L$, 
6(c) \epk \, vs. $L$)  described above are plotted. Besides, Fig.(6d-6f) present three Hubble 
diagrams obtained by varying the normalization constant in the variability relations. Notice that 
Fig.(6e) combines the SWIFT and Schaefer 69 GRBs data. These figures summarize the whole process 
of testing the consistency of the GX model (for several values of the parameters $k$ and $\Omega_m$) 
with SWIFT GRBs data, as shown in Tables \ref{table1}-\ref{table9}. The attentive reader should bear 
in mind that only three of those calibration processes are illustrated in Figure-6.  
Fig.(6d-6f) reveal also the sensitivity of the 
calibration procedure for these definitions of the GRBs variability
on the normalization constant. The combined HD in Fig(6e) appears to be
still consistent with the GX cosmological model, but it is also clear that the SWIFT GRBs data 
distort the better delineate HD of the Schaefer's 69 GRBs sample. This suggests that more data
of high redshift GRBs detected by SWIFT would be needed to clarify the viability of using them as insight 
into GRBs precision cosmology. In this respect, perhaps the use of the SWIFT 77 GRBs sample with 
measured redshifts collected by Butler et al. (2007)  may be of some help. That sample upon
arrival of other new datasets will be used in future studies.

\begin{figure*}[hbt]
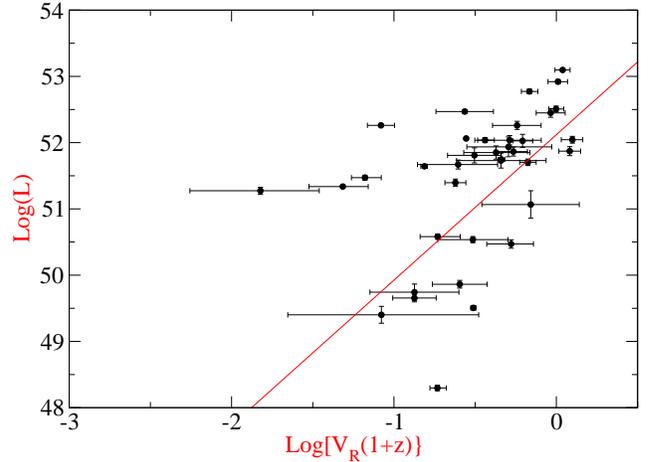
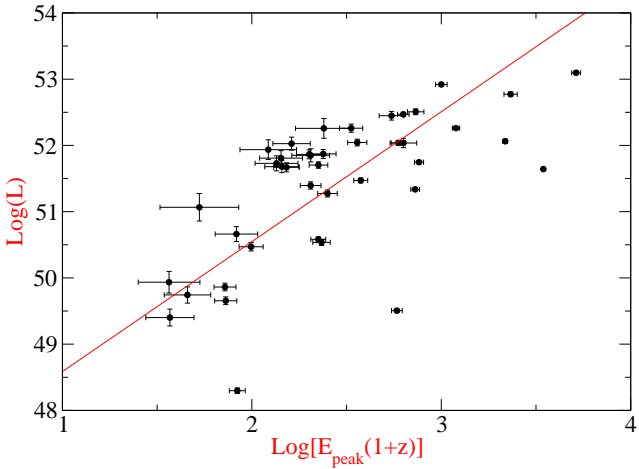
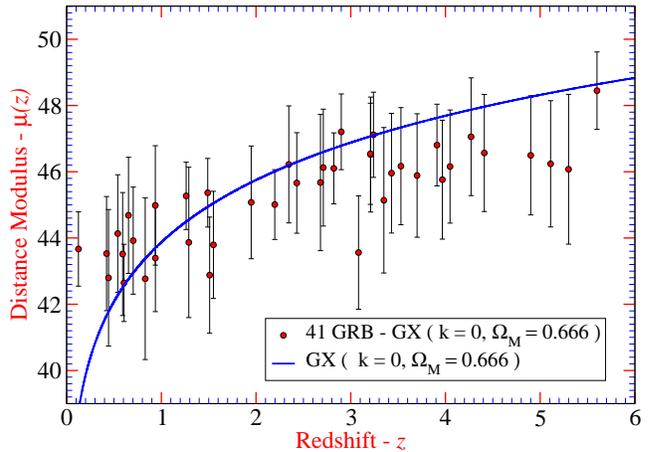
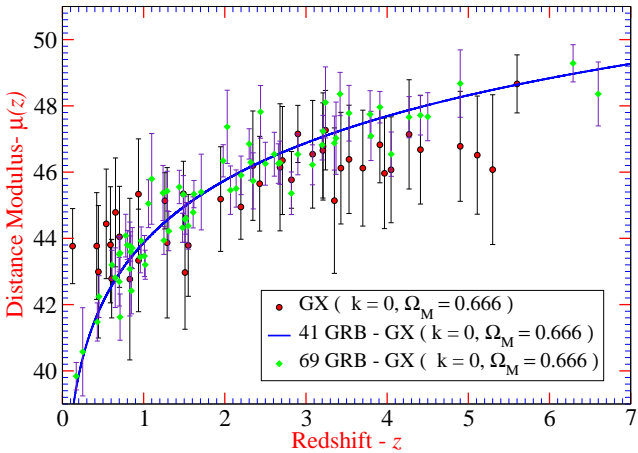
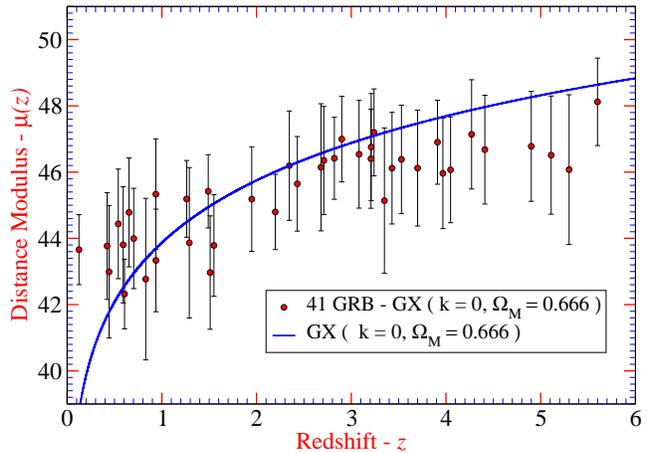
 
{\vspace{0.75truecm}
\subfigure[\label{swift1} Variability $V_{\rm LP}$ vs. $L$ ]{ 
\includegraphics[scale=0.35]{Figure_Swift_M2_Vlp.eps}} 
\hspace{0.25cm} 
\subfigure[\label{swift2} Variability $V_{\rm R}$ vs. $L$]{ 
\includegraphics[scale=0.35]{Figure_Swift_M2_Vr.eps}} }
\vspace{0.75truecm}{
\subfigure[\label{swift3} \epk \, vs. $L$ ]{ 
\includegraphics[scale=0.35]{Figure_Swift_M3.eps}} 
\hspace{0.25cm} 
\subfigure[\label{swift4} Hubble diagram ]{ 
\includegraphics[scale=0.35]{Figure_Swift_Vr.eq.Vlp=0.02.eps}} }
\vspace{0.85cm}{ 
\subfigure[\label{swift5} HD for $V_{\rm R}$, $V_{\rm LP}$ with equal normalization 
constant 0.05]{ 
\includegraphics[scale=0.35]{Figure_Swift_Vr.eq.Vlp=0.05.eps}} 
\hspace{0.25cm}  
\subfigure[\label{swift6}  HD for $V_{\rm R}$, $V_{\rm LP}$ with different normalization 
constant]{ 
\includegraphics[scale=0.35]{Figure_Swift_Vr.ne.Vlp.eps}}   } 
\caption{\label{swift-calibrations} COLOR-ONLINE Results of the calibration procedure
for the SWIFT satellite 41 GRBs with redshift firmly determined (Rizzuto et al. 2007). 
Notice that all relations were corrected to the rest frame of the GRB and also 
by using the luminosity best-fit line obtained from the nonlinear regression method. 
(a) Variability $V_{\rm LP}$ for 13 GRBs vs. isotropic luminosity.
(b) Variability $V_{\rm R}$ for 37 GRBs vs. isotropic luminosity. 
(c) \epk\, values for 41 GRBs vs. peak isotropic luminosity. 
(d) Hubble diagram (HD) from above relations with $V_{\rm R} = V_{\rm LP}$ normalization 
constant 0.02. 
(e) HD combining 41 SWIFT and 69 Schaefer's GRBs samples, using $V_{\rm R}$, $V_{\rm LP}$ 
variabilities with equal normalization constant 0.05. 
(f) $V_{\rm R}$, $V_{\rm LP}$ variabilities with different normalization constant} 
\end{figure*}

} 

\section{Conclusion} 

We have used the approach by Schaefer (2007) to obtain the gamma-ray bursts 
Hubble diagram for the cosmological GX models with dark energy. We have shown 
their fit to the observational data of GRBs with redshift up to $z \sim 6$, and 
have obtained the best fit values for the parameter $\Omega_m$ defining the models, 
see Table \ref{chi2}; e.g. the best fits for GRB69 correspond to $\Omega_m = 0.64 \pm 0.005$,
for SN+RG $\Omega_m = 0.508 \pm 0.004$.

Since GX-models predict the current value of dark energy defining from SN1a data, 
this fact is remarkable in the context of degeneracy between the cosmological 
parameters.  

\begin{table}[tph]
\caption{ \label{chi2} The confidence levels for the used datasets separately and together, with
corresponding values of $\Omega_m$; all cases are for $k=0$.}
\begin{center}
\begin{tabular}{l c c}
\textbf{Sample} & $\chi ^{2}$ & $\Omega _{m}$ \\ \hline
SN244+RG20 			  & 1.23 & 0.508\\ 
GRB69 		 			  & 1.04 & 0.64 \\ 
GRB41 		 			  & 1.41 & 0.66 \\ 
SN244+RG20+GRB69  & 1.58 & 0.66 \\ 
GRB69+GRB41 		  & 1.16 & 0.66 \\ 
GRBAll+SN244+RG20 & 1.56 & 0.66 \\ \hline
\end{tabular}
\end{center}
\end{table}

The key issue in such an analysis is up to which extent the 
gamma-ray bursts can be used as standard candles, i. e. in revealing of the 
genuine scatter in the 
empirical relations used for obtaining  the Hubble diagram. Given the lack in 
the understanding 
of the physics of those relations, handling the new data (Sato et al 2007;
Sakamoto et al 2007; Rizzuto et al 2007) 
must need a particular care. For example, recently Campana et al 
(2007) claimed 
weakening of the Ghirlanda relation ($\chi^2$ up to 2 or 3) using a sample of 5 
bursts observed 
by SWIFT. Even assuming that this result is correct, it 
will influence our evaluations negligibly ($\chi^2=1.040$ instead of 1.037). However, 
Ghirlanda et al 
(2007) reconsidered the same SWIFT sample and showed that, that relation survives 
practically unmodified. Recently also Cabrera et al (2007) have reported the correspondence
of a sample of SWIFT data with those of other satellites and, hence, absence of outliers.

In sum, the preliminary character of testing of cosmological models 
via the gamma-ray burst Hubble diagram indeed remains until the understanding
of the nature of the used empirical relations and availability of more observational data.

We thank the referee for many helpful comments. The last three authors are partially supported by INTAS.

\begin{table}[tph]
\centering\ \ 
\caption{\label{table1} Calibration relations for the GX cosmological model with parameters $k$ and 
$\Omega _{m}$ as indicated. $A$ and $B$ represent the linear regression intersept and 
slope as defined in the text, while $\sigma _{A}$ and $\sigma _{B}$ represent their
respective errors, obtained through the linear regression after using the SLOPES method.
Here $L$ is the luminosity, $\tlag$ is the time lag, $V$ is the variability, $E_{peak}$ 
is the peak (or maximum) energy, $\eg$ is the corresponding energy in gamma-rays, and 
$\tau_{RT}$ is the rise-time.}
\begin{tabular}{l c c c c}
Luminosity Relation & A & B & $\sigma _{A}$ & $\sigma _{B}$ \\ 
$k=0,\,\, \Omega_{m}=0.6666$&       &            &        
                    \\ \hline
$\tau_{\text{lag}}-L$        & 52.2 & -1.08  & \ 0.0724 &  0.12 \\
$V-L$ 											 & 52.4 & \ 1.76 &  0.102   & \ 0.247 \\ 
\epk-$L$ 										 & 52.2 & \ 1.71 & \ 0.0548 & \ 0.112 \\
\epk-E$_{\gamma}$ 					 & 50.5 & \ 1.62 & \ 0.0525 & \ 0.119 \\ 
$\tau_{\text{\tiny RT}}-L$ 	 & 52.5 & -1.25  & \ 0.0691 & \ 0.122 \\ 
\hline
\end{tabular}
\end{table}

\begin{table}[tph]
\centering\ \ 
\caption{\label{table2} Calibration relations for the GX cosmological model with parameters $k$ and 
$\Omega _{m}$ as indicated.}
\begin{tabular}{l c c c c}
Luminosity Relation & A & B & $\sigma _{A}$ & $\sigma _{B}$ \\ 
$k=0,\,\,\,\Omega _{m}=0.633$ &       &          &          &\\
\hline
$\tau_{\text{lag}}-L$ 		 &   52.2   &  -1.09   &  0.0747 &   0.123 \\ 
$V-L$ 										 &   52.4   & \ 1.71   & \ 0.101 &   0.243 \\ 
\epk-$L$ 									 &   52.1   & \ 1.67   &  0.0539 &   0.112 \\ 
\epk-E$_{\gamma}$     		 &   50.4   & \ 1.6    &  0.0538 &   0.121 \\ 
$\tau_{\text{\tiny RT}}-L$ &   52.4   &  -1.22   &  0.0676 & \ 0.12 \\ \hline
\end{tabular}
\end{table}

\begin{table}[tph]
\centering\ \ 
\caption{Calibration relations for the GX cosmological model with parameters $k$ and 
$\Omega _{m}$ as indicated.}
\label{table3}
\begin{tabular}{l c c c c }
Luminosity Relation & A & B & $\sigma _{A}$ & $\sigma _{B}$ \\ 
$k=0, \,\,\,\Omega _{m}=0.645$ & & & \\ \hline
$\tau_{\text{lag}}-L$      &   52.2  &     -1.09 & \ 0.0733 & \ 0.121 \\ 
$V-L$ 										 &   52.5  &  \  1.78  &  0.103   & \ 0.248 \\ 
\epk-$L$ 									 &   52.2  &  \  1.73  & \ 0.0552 & \ 0.112 \\ 
\epk-E$_{\gamma}$ 				 &   50.5  &  \  1.63  & \ 0.0522 & \ 0.119 \\ 
$\tau_{\text{\tiny RT}}-L$ &   52.5  &     -1.26 & \ 0.07   & \ 0.123 \\ \hline
\end{tabular}
\end{table}

\begin{table}[tph]
\centering\ \ 
\caption{\label{table4} Calibration relations for the GX cosmological model with parameters $k$ and 
$\Omega _{m}$ as indicated.}
\begin{tabular}{l c c c c }
Luminosity Relation & A & B & $\sigma _{A}$ & $\sigma _{B}$ \\ 
$k=0, \,\,\,\Omega _{m}=0.76$& & & &\\ \hline
$\tau_{\text{lag}}-L$ 	   &  52.1 &   -1.05  &   0.0706 &   0.118 \\ 
$V-L$ 									   &  52.4 &  \ 1.71  & \ 0.101  &   0.243 \\ 
\epk-$L$ 								   &  52.1 &  \  1.67 &   0.0523 &   0.111 \\ 
\epk-E$_{\gamma}$ 			   &  50.4 &  \  1.6  &   0.0538 &   0.121 \\ 
$\tau_{\text{\tiny RT}}-L$ &  52.4 &   -1.22  &   0.0678 & \ 0.12 \\ \hline
\end{tabular}
\end{table}

\begin{table}[tph]
\centering\ \ 
\caption{\label{table5} Calibration relations for the GX cosmological model with parameters $k$ and 
$\Omega _{m}$ as indicated.}
\begin{tabular}{l c c c c}
Luminosity Relation & A & B & $\sigma _{A}$ & $\sigma _{B}$ \\
$k=0, \,\,\,\Omega _{sep}=2/3$& & & & \\ \hline
$\tau_{\text{lag}}-L$ 		   &  \ 52.2  &  -1.08   &   0.0723 & \ 0.12 \\ 
$V-L$ 										   &  \ 52.4  & \ 1.76   & \ 0.102  &   0.247 \\ 
\epk-$L$ 										 & {\ 52.2} & \ 1.71   &   0.0548 &   0.112 \\ 
\epk-E$_{\gamma}$ 					 & {\ 50.5} & \ 1.62   &   0.0525 &   0.119 \\ 
$\tau_{\text{\tiny RT}}-L$   &  \ 52.5  &  -1.25   &   0.0691 &   0.122 \\ \hline
\end{tabular}
\end{table}

\begin{table}[tph]
\centering\ \ 
\caption{\label{table6} Calibration relations for the GX cosmological model with parameters $k$ and 
$\Omega _{m}$ as indicated.}
\begin{tabular}{l c c c c}
Luminosity Relation & A & B & $\sigma _{A}$ & $\sigma _{B}$ \\
$k=-1, \,\,\,\Omega _{m}=0.566$& & & &\\ \hline
$\tau_{\text{lag}}-L$ 		 & {\ 52.3} & {\ -1.11} & \ 0.0755 & \ 0.123 \\ 
$V-L$ 										 & {\ 52.5} & {\ 1.82} 	& 0.104 	 & \ 0.25 \\ 
\epk-$L$ 									 & {\ 52.3} & {\ 1.77} 	& \ 0.0562 & \ 0.112 \\ 
\epk-E$_{\gamma}$ 				 & {\ 50.6} & {\ 1.66} 	& \ 0.0516 & \ 0.125 \\ 
$\tau_{\text{\tiny RT}}-L$ & {\ 52.6} & {\ -1.29} & \ 0.0724 & \ 0.125 \\ \hline
\end{tabular}
\end{table}

\begin{table}[tph]
\centering\ \ 
\caption{\label{table7} Calibration relations for the GX cosmological model with parameters $k$ and 
$\Omega _{m}$ as indicated.}
\begin{tabular}{l c c c c}
Luminosity Relation& A & B & $\sigma _{A}$ & $\sigma _{B}$ \\ 
$k=-1, \,\,\,\Omega _{m}=0.578$& & & & \\ \hline
$\tau_{\text{lag}}-L$ 		 & {\ 52.3} & {\ -1.1}  & \ 0.0746 & \ 0.122 \\ 
$V-L$ 										 & {\ 52.5} & {\ 1.81}  & 0.103    & \ 0.249 \\
\epk-$L$ 									 & {\ 52.2} & {\ 1.75}  & \ 0.0558 & \ 0.112 \\ 
\epk-E$_{\gamma}$ 				 & {\ 50.6} & {\ 1.65}  & \ 0.0521 & \ 0.121 \\ 
$\tau_{\text{\tiny RT}}-L$ & {\ 52.6} & {\ -1.28} & \ 0.0713 & \ 0.124 \\ \hline
\end{tabular}
\end{table}

\begin{table}[tph]
\centering\ \ 
\caption{\label{table8} Calibration relations for the GX cosmological model with parameters $k$ and 
$\Omega _{m}$ as indicated.}
\begin{tabular}{l c c c c}
Luminosity Relation & A & B & $\sigma _{A}$ & $\sigma _{B}$ \\
$k=-1,\,\,\,\Omega_{m}=0.605$ & & & & \\ \hline
$\tau_{\text{lag}}-L$ 		 & {\ 52.3} & {\ -1.09} &  \ 0.0732 & \ 0.121 \\ 
$V-L$ 										 & {\ 52.5} & {\ 1.78}  &     0.103 & \ 0.248 \\ 
\epk-$L$ 									 & {\ 52.2} & {\ 1.73}  &  \ 0.0551 & \ 0.112 \\ 
\epk-E$_{\gamma}$ 				 & {\ 50.5} & {\ 1.63}  &  \ 0.0522 & \ 0.119 \\ 
$\tau_{\text{\tiny RT}}-L$ & {\ 52.5} & {\ -1.26} &  \ 0.0698 & \ 0.123 \\ \hline
\end{tabular}
\end{table} 

\begin{table}[hbt]
\centering\ \ 
\caption{ \label{table9} Calibration relations of the SWIFT satellite GRBs data for the GX cosmological 
model with parameters $k$ and $\Omega _{m}$ as indicated. Several values of the normalization 
constant were used in an attempt to understand the sensitivity of the resulting HD to such a 
parameter (see Fig.\ref{swift-calibrations}d, \ref{swift-calibrations}f). Also the systematic error is provided.}
\begin{tabular}{l c c c c c}
Luminosity Relation& A & B & $\sigma _{A}$ & $\sigma _{B}$ & $\sigma_{sys}$ \\
$k=0,\,\,\,\Omega _{m}=0.666$ & & & & & \\ \hline
$V-L$(Vr=0.05)   		& \ 49.902 &  1.98   & \ 0.487 & \ 0.448 &  \ 0.729 \\ 
$V-L$(Vr=0.02)   		& \ 49.902 &  1.98   & \ 0.487 & \ 0.966 & \ 0.966 \\ 
$V-L$(Vlp=0.05)  		& \ 52.744 & \ 1.634 & \ 0.222 &  0.3    & 0.3 \\ 
$V-L$(Vlp=0.005) 		&  51.11   & \ 1.634 & \ 0.443 & \ 0.352 & \ 0.352 \\ 
\epk-$L$(Band=1MeV) & 52.5     & \ 1.961 &  0.19   & \ 0.805 & \ 0.805 \\ \hline
\end{tabular}
\end{table}

\end{document}